# TRANSMISSION ELECTRON MICROSCOPY STUDIES ON RF SPUTTERED COPPER FERRITE THIN FILMS

Prasanna D. Kulkarni and Shiva Prasad
Department of Physics
IIT Bombay, Mumbai 400076

Indradev Samajdar
Metallurgical Engineering & Materials Science
IIT Bombay, Mumbai 400076

N. Venkataramani
ACRE IIT Bombay
Powai, Mumbai 400076

R. Krishnan
Laboratoire de Magnetisme et d'optique de Versailles,
CNRS, 78935 Versailles, France

ABSTRACT
Copper ferrite thin films were rf sputtered at a power of 50W. The as deposited films were annealed in air at 800°C and slow cooled. The transmission electron microscope (TEM) studies were carried out on as deposited as well as on slow cooled film. Significantly larger defect concentration, including stacking faults, was observed in 50W as deposited films than the films deposited at a higher rf power of 200W. The film annealed at 800°C and then slow cooled showed an unusual grain growth upto 180nm for a film thickness of ~240nm. These grains showed Kikuchi pattern.

INTRODUCTION
It has been observed that changing power significantly alters the microstructure and magnetic properties of the rf sputtered films. In the case of hexagonal strontium ferrite films, the films are oriented when deposited at a low rf power[1]. Recently, our group has carried out extensive studies on copper ferrite films[2]. The copper ferrite films have been deposited at different rf powers and subjected to ex-situ thermal treatment. Copper ferrite exhibits two different phases. A high temperature cubic phase and a room temperature tetragonal phase. The high temperature phase can be stabilised at room temperature after the films are annealed and quenched. The detailed structural and magnetic properties of the 200W deposited copper ferrite films have been reported earlier[2]. The as deposited films are cubic. For the slow cooled films, a phase transformation from cubic to tetragonal has been observed on heating the film. When copper ferrite thin films are deposited at a lower rf power of 50W, a strong ($\ell\ell\ell$) orientation is observed for the quenched films. These orientation features and magnetic properties of the quenched 50W copper ferrite films are reported earlier[3]. In this paper, the detailed TEM studies carried out to investigate the microstructure and magnetic properties of the 50W as deposited and slow cooled films are reported.

EXPERIMENTAL
The copper ferrite films are deposited using the Leybold Z400 rf sputtering system. The films are deposited on amorphous quartz as well as silicon (111) substrates. No heating or cooling was carried out during sputtering. The rf power employed during the deposition is 50W. The film thickness is ~240nm. The as deposited film is ex-situ annealed, at 800°C, followed by slow cooling. The structure of the as deposited and slow cooled films are studied using X-ray diffraction (Phillips X-pert system) and the microstructure by Transmission Electron Microscope (Phillips CM200). The samples for TEM are prepared by chemical etching of silicon in a 3:1 solution of HNO₃ and HF. Magnetization studies are carried out using Vibrating Sample Magnetometer (VSM), up to a field of 0.7T.

RESULTS AND DISCUSSION

Fig.1(a) shows the XRD pattern for the as deposited copper ferrite film deposited at 50W rf power. A broad hump at a 2θ value of 35 degree is observed in this case. A 100% intensity peak of copper ferrite is centred around this 2θ value. This has been reported as a (311) peak in the JCPDS data for bulk copper ferrite.

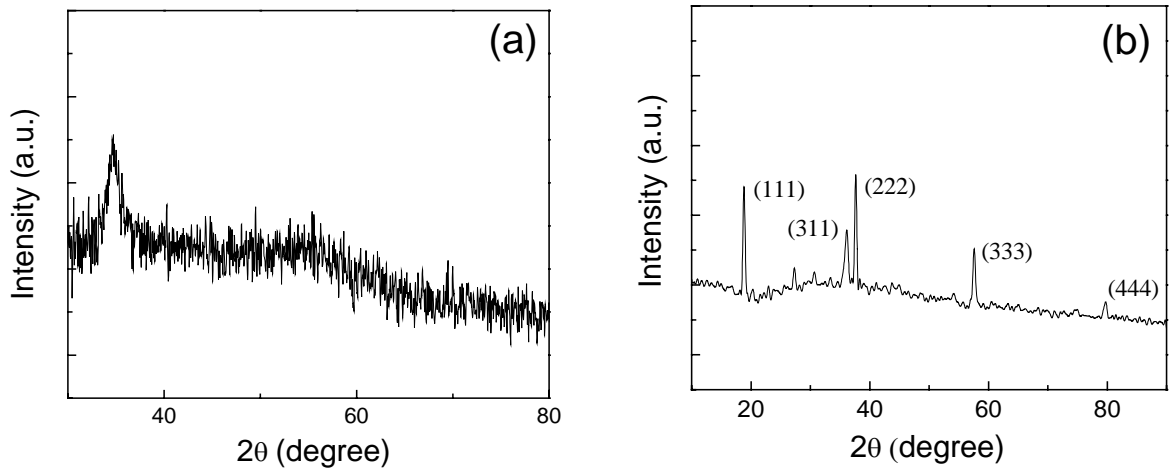

Fig. 1. XRD patterns of 50W copper ferrite films, (a) As deposited (b) 800°C Slow cooled.

The as deposited film has been annealed in air at 800°C and slow cooled. The XRD pattern of the slow cooled film is shown in the Fig. 1(b). The unit cell of the tetragonal phase of copper ferrite has been transposed to a pseudo cubic unit cell[2]. Miller indices of the planes indexed to the XRD pattern are with respect to the transposed unit cell. This is done to facilitate the comparison with the cubic phase of copper ferrite. All the peaks observed in the XRD pattern could be indexed to the copper ferrite spinel structure. The (111), (222), (333), (444) peaks are present along with a (311) peak. The (222) peak which is about 8% intense peak in the bulk copper ferrite is a 100% intensity peak in the slow cooled film. This shows that the slow cooled film has a strong (111) orientation, which is the plane containing the highest packing density for the spinel structure. These orientation features are not observed in the 200W deposited films[2], but are seen in the 50W quenched film. The usual splitting of (333) and (511) peaks, which is a signature of the presence of tetragonal phase is not seen in the XRD pattern of slow cooled 50W film. This may be due to the strong orientation present in the film. Hence, the phase of the slow cooled film in the low power case is not conclusive from XRD pattern alone.

The Transmission Electron Microscope (TEM) studies are carried out on the 50W as deposited and the slow cooled films. Fig.2(a) shows the TEM bright field image and the corresponding diffraction pattern for the as deposited copper ferrite film. In the bright field image, large defect concentration is observed, along with a presence (at higher magnification) of stacking faults. The grains with similar orientations are clusterd, as confirmed by conical dark field imaging, with the cluster sizes in the range of 10-25nm.

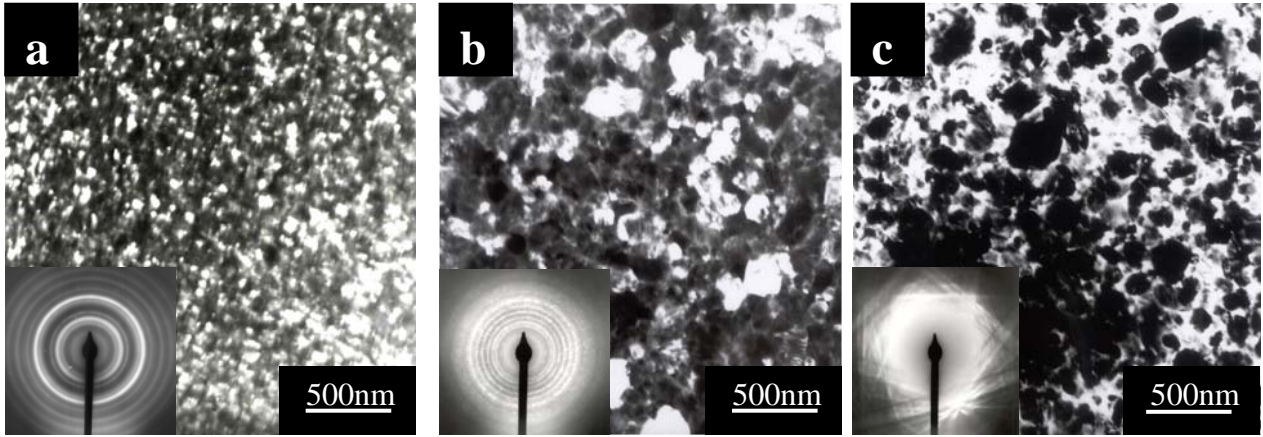

Fig. 2. TEM images for 50W copper ferrite films, (a) As deposited, (b) 800°C Slow cooled (bright field image), (c) 800°C Slow cooled (conical dark field image).

The bright field image for the slow cooled film and the corresponding diffraction pattern is shown in Fig.2(b). The crystallites within 5 degree misorientation or falling within near Bragg, are considered as a grain. The grain sizes are identified by conical dark field imaging in TEM. Fig. 2(c) shows the dark field image of the slow cooled film. The grain sizes lie in the range 10-180nm. Clear signature of abnormal grain growth is observed in the film. The large defect concentration and clustering of the grains of similar orientation is not observed in the as deposited films of 200W[2]. Additionally the grain growth in 200W deposited and then annealed at 800°C are restricted at ~40nm. This has been attributed to pinning by the low angle grain boundaries[2]. For the same growth mechanism, a transition from normal to abnormal grain growth can be caused by pinning (either through second phase or through higher low angle boundary concentration) or relative differences in the concentration of defects[4]. In view of the fact that defet concentration in as deposited 50W film is larger than 200W film, the growth seems to be because of relative difference in concentration of defects. Both 50W and 200W films go through a phase transformation while annealing at 800°C. The 200W film, however, does not show the extent of defect elimination, as observed in the case of 50W film through Kikuchi. The Kikuchi pattern observed for a 50W slow cooled film is shown in the inset of Fig.2(c). To the best knowledge of authors, this is the first observation of Kikuchi in ferrite thin films. Kikuchi formation needs a subtle balance of elastic and inelastic scattering[5]. The later, should be stronger in nano crystalline ferrite films due to high defect concentration and should not allow presence of Kikuchi. In the 50W annealed material, however, even from convergent spots corresponding to lower grain sizes (~20nm) Kikuchis are clearly visible in TEM. This is an indirect indication of the elimination of defects in the 50W slow cooled films.

Fig 2(a) shows the diffraction pattern for the as deposited film. The diffraction pattern of the as deposited film could be indexed to the cubic phase of copper ferrite. Inset of Fig.2(b) shows the diffraction pattern for the slow cooled film. The splitting[2] of the rings corresponding to the same $h^2+k^2+l^2$ of a plane is not observed in the diffraction pattern. Such a splitting of the rings is clearly observed in 200W slow cooled film[2]. This may be due to the orientation present in the film. However, TEM results when taken with the magnetisation data (to be discussed next) leads to the inference that the slow cooled film is in tetragonal phase.

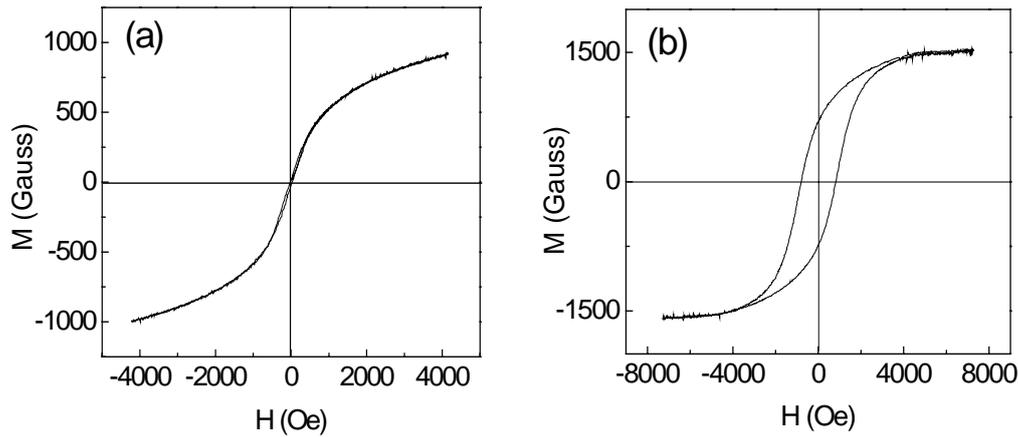

Fig.3 Parallel MH loops for 50W copper ferrite films, (a) As deposited, (b) 800°C Slow cooled.

The magnetic measurements have been done on the as deposited as well as slow cooled 50W copper ferrite films. Fig.3(a) shows the parallel MH loop for the as deposited 50W film. The saturation magnetization calculated from the parallel MH loops is ~750 Gauss. This value is much lower than the saturation magnetization of ~1100 Gauss in the 200W as deposited films[2]. The magnetic properties of the slow cooled 50W film are similar to the slow cooled 200W film. Fig. 3(b) shows the parallel MH loop for the 50W slow cooled film. The saturation magnetization calculated from the parallel MH loop is ~1500 Gauss. The saturation magnetization in case of 200W deposited slow cooled flims is ~1600 Gauss. The smaller value of saturation magnetization in 50W slow cooled film than in case of 200W slow cooled film, is not well understood. The mangetization realised in the 50W and 200W slow cooled films may not be solely due to grain sizes.
In any case, the saturation mangetization value of ~1500 Gauss for 50W slow cooled film is close to the bulk value (~1700 Gauss) for tetragonal copper ferrite. This confirms that the slow cooled films are in tetragonal phase of copper ferrite.

CONCLUSIONS
The present study shows the rf power employed during sputtering alters the microstructure of the copper ferrite thin films significantly. At a lower rf power the defect concentration increases along with orientation along the [111] direction. Due to the crystalline orientation, the specific peaks identifying the tetragonal phase are not observed in XRD and TEM data. The magnetization value confirms the presence of the tetragonal phase. The slow cooled films show the abnormal grain growth and Kikuchi pattern.

ACKNOWLEDGEMENT
The author Prasanna D. Kulkarni thanks the CSIR, India for financial support. Two of the authors Shiva Prasad and N. Venkataramani thank the IIT Bombay heritage fund for the financial support to attend the conference.

REFERENCES

[1] B. Acharya, S.N. Piramanayagam, A. Ajan, S.N. Shringi, S. Prasad, N. Venkataramani, R. Krishnan, S.D. Kulkarni, S.K. Date, "Oriented strontium ferrite films sputtered onto Si(111)," *Journal of Magnetism and Magnetic Materials* 140-144 (1995) 723-724.

[2] M.Desai, S. Prasad, N. Venkataramani, I. Samajdar, A.K. Nigam, R. Krishnan, "Annealing induced structural change in sputter deposited copper ferrite thin films and its impact on magnetic properties," *Journal of Applied Physics*, 91[4] 2220-27 (2002).

[3] P. Kulkarni, M. Desai, N. Venkataramani, S. Prasad, R. Krishnan, "Low power RF sputter deposition of oriented copper ferrite films," *Journal of Magnetism and Magnetic Materials* 272-276 (2004) e793-e794.

[4] F.J.Humphreys and M. Hatherly, "Abnormal grain growth"; pp. 314-325 in *Recrystallization and Related Annealing Phenomena*, 1st Edition, Pergamon, New York, 1995.

[5] P.J. Goodhew, J. Humphreys, R. Beanland, "Kikuchi line patterns"; pp. 58-61 in *Electron Microscopy and Analysis*, 3rd ed., Taylor & Francis, London, 2000.